# Electronic Strong Coupling of Gas-Phase Molecular Iodine

Jane C. Nelson, Trevor H. Wright, Neo Lin, Madeline Rohde, Marissa L. Weichman*

Department of Chemistry, Princeton University, Princeton, New Jersey 08544, United States

**ABSTRACT:** Molecular polaritons, hybrid light-matter states formed from the strong coupling of molecular transitions and discrete photonic modes, are a compelling platform for optical control of chemical reactivity. Despite the origins of the field of polaritonics in atomic gases, strong coupling of molecular gases remains underexplored. The pristine, solvent-free gas-phase environment may prove ideal for gaining mechanistic understanding of molecular behavior under strong light-matter coupling. In this work, we achieve electronic strong coupling of the B−X, $\nu_1$ = 0→32, $J$ = 53→52 and B−X, $\nu_1$ = 0→34, $J$ = 103→102 rovibronic transitions of gas-phase iodine ($I_2$) lying near 532.2 nm. We access a range of coupling strengths and detuning conditions with fine control over molecular number density and cavity length stabilization. This effort represents the first demonstration of electronic polaritons in a molecular gas and opens a new platform for polariton photochemistry and photophysics.

## 1. INTRODUCTION

Precisely tailored light-matter interactions are central to many topics at the frontiers of physical chemistry, including ultrafast and precision spectroscopy, energy transfer, photochemistry, and photocatalysis. Strong light-matter coupling between optical cavity modes and molecular transitions has emerged as a compelling new playground for optical control of reactivity.[1–3] Polaritons are hybrid light-matter states that arise when photonic and material transitions are strongly coupled. While polaritons have been demonstrated in a variety of media including atoms,[4,5] semiconductor quantum wells and quantum dots,[6–8] and other platforms,[9] the potential for strong cavity-coupling to alter chemical reactions has driven much of the last decade of study.[1–3,10–12] Both vibrational strong coupling (VSC) arising from cavity coupling to molecular vibrations in the infrared and electronic strong coupling (ESC) of molecular electronic transitions at optical wavelengths are of interest for new photonic control schemes.[2,3,9–14] In our previous work, we made the first demonstration of gas-phase molecular VSC by cavity-coupling rovibrational transitions of methane.[15,16] Here, we achieve ESC in the rovibronic transitions of gas-phase molecular iodine ($I_2$) under similar conditions in a centimeter-scale Fabry-Pérot optical cavity.

The field of polariton chemistry remains in search of a comprehensive, predictive theory for how molecules undergo photochemistry under ESC.[3,10–12] This question was first explored by Hutchison *et al.* in 2012,[17] who reported slowing of a photoswitching reaction under strong cavity coupling to an electronic transition of the product. At the time, the authors ascribed this result to cavity modification of the excited state potential energy surfaces. Over the past decade, a handful of other altered photochemical reaction rates under ESC have been reported,[18–20] alongside null results[21,22] and questions about reproducibility.[23–25] Altered energy transfer and exciton transport in molecular films and solid-state materials under ESC has also been of great interest in the field.[26–31] ESC is most commonly achieved with molecular thin films embedded in planar Fabry-Pérot cavities with lengths on the order of hundreds of nanometers.[3,11,12] Two demonstrations have also been made with solution-phase molecules in similarly compact cavities.[32,33] Condensed-phase molecules feature broad, unresolved electronic bands, requiring the use of very short cavities with correspondingly large free spectral ranges (*FSR*s) to avoid spectral congestion. In addition, very dense molecular films are often used to reach the collective ESC regime, leading to aggregation and excimer formation. These many-body interactions can make analysis of excited-state molecular dynamics under ESC difficult[34–36] and also complicate comparison with theory, which has largely focused on single-molecule photochemistry under ESC.[37–42] Our aim here is to extend molecular ESC into the gas phase where intermolecular interactions are minimized, enabling simpler studies of photochemistry under ESC with closer connection to theoretical work.

Strong light-matter coupling can be understood within either a quantum optics or classical optics description. In the quantum optics treatment, coherent energy exchange between the material and photonic sub-systems outcompetes their individual dissipative pathways.[42–44] These dynamics manifest as a frequency-domain splitting of the coupled cavity mode by a Rabi frequency ($\Omega_R$) that exceeds the full-width at half-maximum (fwhm) linewidths of the original material (2γ fwhm linewidth) and photonic (2κ fwhm linewidth) constituents and can therefore be observed in the cavity transmission spectrum (Figure 1a). The two new split modes represent the lower and upper polariton states. In the collective strong coupling regime, $N$ molecules are packed into the cavity and $\Omega_R$ scales as $(N/V)^{1/2}$, where $V$ is the cavity mode volume.[45] In addition, $N$–1 dark states arise from optically-forbidden linear combinations of molecular excitations and remain at the uncoupled molecular energy (Figure 1b). Some uncertainty persists around the role of dark states in cavity-mediated reactivity and photophysics, though it is likely that the dark states localize and behave like uncoupled molecules in the presence of disorder.[46,47]

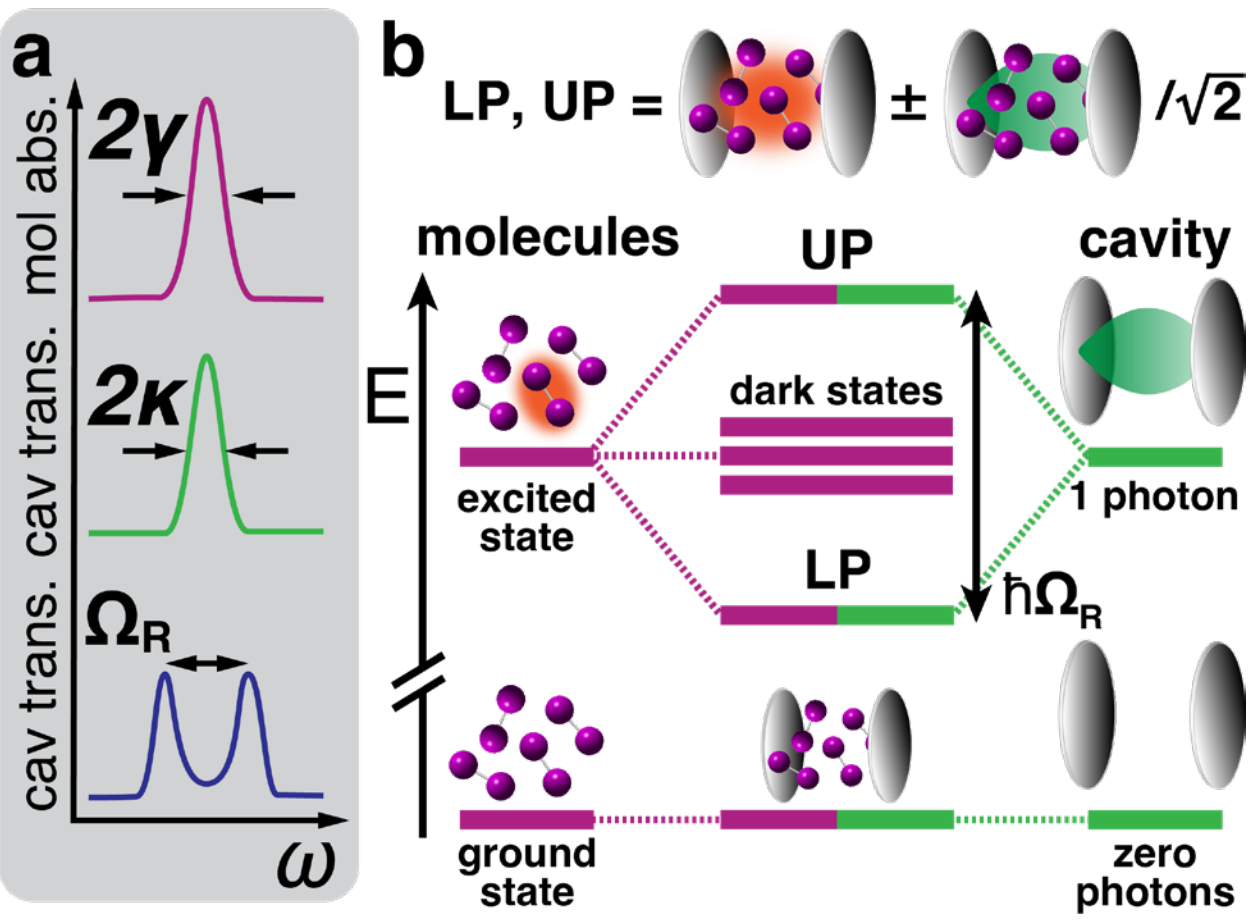

**Figure 1. (a)** Spectroscopic signatures of a polaritonic system. Under strong light-matter coupling, the characteristic Rabi splitting ($\Omega_R$) in the cavity transmission spectrum exceeds the full-width at half-maximum (fwhm) linewidths of the original molecular (magenta, 2γ fwhm linewidth) and cavity (green, 2κ fwhm linewidth) modes. **(b)** Within collective quantum optics, the hybridization of $N$ molecular excitations and a single quantized cavity mode gives rise to the lower and upper polariton (LP and UP) states and $N$–1 dark states. The polaritons represent superpositions of molecular and photonic excitations.

While the quantum optics picture of polariton formation is intuitive, it is of debatable practical relevance to the disordered, large-$N$ systems commonly realized in experiments.[48,49] Classical optical cavity physics is well-known to capture many features observed in both linear and nonlinear cavity spectra under strong coupling.[50–53] As in our previous work,[15,16,54] we model our strongly-coupled system as a Fabry-Pérot cavity composed of two identical mirrors and filled with a homogeneous dielectric intracavity medium. The fractional intensity of light transmitted through such a structure is given by[50,51]

$$\frac{I_T(\nu)}{I_0} = \frac{T^2 e^{-\alpha(\nu)L}}{1+R^2 e^{-2\alpha(\nu)L} - 2Re^{-\alpha(\nu)L}\cos[\delta(\nu)]} \qquad (1)$$

where $L$ is the cavity length, $R$ and $T$ are the reflection and transmission intensity coefficients of the cavity mirrors, $\alpha(\nu)$ is the frequency-dependent absorption coefficient of the intracavity medium, $\delta(\nu) = 4\pi L n(\nu)\nu/c$ is the round-trip phase accrued by light traveling in the cavity, $n(\nu)$ is refractive index of the intracavity medium, and $c$ is the speed of light. The absorption coefficient is given by $\alpha(\nu) = \sigma(\nu) \cdot N/V$, where $\sigma(\nu)$ is the absorption cross section. $n(\nu)$ can be derived from $\alpha(\nu)$ via the Kramers-Kronig relation. Peaks in the cavity transmission spectrum arise from constructive interference whenever the round-trip phase $\delta(\nu)$ is equal to an integer multiple of 2π, minimizing the denominator of eq 1. When $n(\nu)$ takes on a constant value $n_0$, transmission maxima are evenly spaced by the cavity free spectral range, $FSR = c/2n_0L$. However, near a strong absorption band of intracavity molecules, $n(\nu)$ takes on a dispersive lineshape and yields additional transmission features whose splitting coincides with the collective Rabi splitting predicted by quantum optics. Using eq 1, we can readily predict the transmission spectra of strongly-coupled cavities with simple classical optics. It remains an open question in the field whether this classical description can explain reports of unexpected chemistry and photophysics of polaritonic molecules, or if the more exotic quantum optics treatment must be invoked.

Here, we demonstrate electronic molecular polaritons in the gas phase. The gas phase is the historic proving ground of chemical physics, where simple benchmark molecules can be studied with clean spectroscopic readout, quantum-state resolution, a high degree of experimental control, and accurate theoretical treatment. We recently reported the first demonstration of molecular gas-phase VSC in methane.[15,16] We now extend these capabilities to ESC by cavity-coupling individual rovibronic transitions of $I_2$ gas. This platform is highly tunable and the open centimeter-scale cavity geometry will permit orthogonal *in situ* spectroscopic access to intracavity molecules. We expect this approach to aid in measurement, interpretation, and mechanistic understanding of any future observations of cavity-modified photophysics and photochemistry.[54]

## 2. EXPERIMENTAL SECTION

Here, we achieve strong coupling of rovibronic transitions of $I_2$ molecules near 532.2 nm in a Fabry-Pérot optical cavity mounted in a home-built gas cell. As we recently predicted,[54] the collective strong coupling regime is accessible in gas-phase $I_2$ at a modest molecular number density near the room-temperature vapor pressure. We deliver $I_2$ vapor to the cavity cell via helium carrier gas and probe the cavity transmission spectrum at different flow rates with continuous-wave (cw) spectroscopy.

We mount the optical cavity used for strong coupling on a commercial fiber bench (Figure 2a), inspired by implementations in prior literature.[55,56] We seal the fiber bench cover in place with epoxy to create a gas cell and evacuate the cell with a roughing pump to a modest base pressure of ~14 torr. We use an inexpensive roughing pump due to the contact with corrosive $I_2$. To deliver $I_2$ to the cell, we pass He carrier gas through a sealed glass bottle containing solid $I_2$ crystals, which are sourced from Sigma-Aldrich and used without further purification. A mass flow controller is used to regulate the He flow from 0 to 100 standard cubic centimeters per min (sccm). At the highest flow rates, the cell reaches a maximum pressure of ~33 torr. Because we rely on the room-temperature vapor pressure of the $I_2$ sample, we control the intracavity [$I_2$] indirectly via the He flow rate. We expect the intracavity [$I_2$] to saturate near the room-temperature vapor pressure[57] of 0.23 torr, corresponding to $N/V \sim 7.4 \times 10^{15}$ cm$^{-3}$. More details on this experimental set-up are provided in Section S1.1 of the Supporting Information (SI).

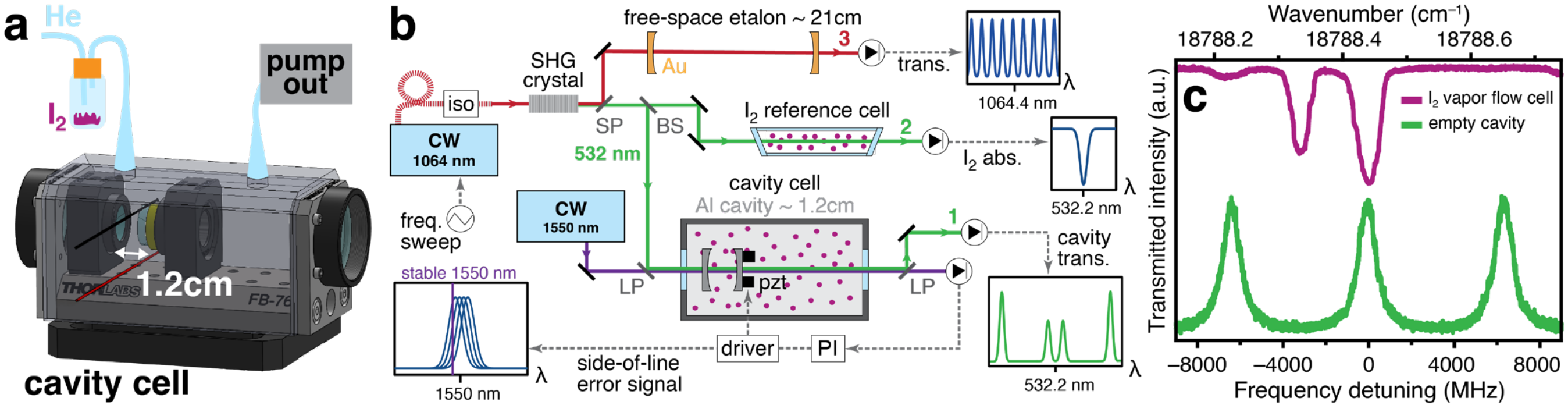

**Figure 2. (a)** Simplified schematic of the cavity cell. Helium carrier gas flows through a bottle containing $I_2$ crystals before entering the cell and exiting to the pump. One cavity mirror is held in a lens tube mount while the other is glued to a ring-shaped piezoelectric chip then glued to its lens tube mount. Both tube mounts are secured to a fiber bench such that the cavity mirrors are concentric and held ~1.2 cm apart. CAD models of commercial optomechanical components (FB-76W, FBS05, SM05L05, and SM1L03) courtesy of Thorlabs, Inc. are used with permission. **(b)** Laser scheme used to record cavity transmission spectra, perform absolute and relative frequency calibration, and stabilize the optical cavity length. 1064 nm light from a distributed feedback diode laser (red) is fed through a fiber-coupled optical isolator (iso), then coupled into free space before frequency doubling in a second-harmonic generation (SHG) crystal to generate green 532 nm light. Following the SHG crystal, light is divided into three arms to acquire cavity transmission spectra (beam 1) and calibration spectra (beams 2 and 3). The residual 1064 nm light is separated with a shortpass (SP) dichroic mirror and directed to a free-space etalon for relative frequency calibration (beam 3). The 532 nm light passes through a homemade Al flat used as a beamsplitter (BS); 82% of the beam is mode-matched into the Al optical cavity (beam 1) while 3% is directed to a commercial $I_2$ vapor cell to serve as an absolute frequency reference (beam 2). We stabilize the cavity length via a side-of-line lock, locking one cavity fringe to a metrology-grade 1550 nm diode laser. The 1550 nm light is coupled in and out of the cavity with longpass (LP) dichroic mirrors. The locking error signal is processed with a proportional-integral (PI) loop filter and fed back onto the cavity piezoelectric chip (pzt). The cavity cell in beam 1 can be replaced with a vapor flow cell to make extracavity $I_2$ measurements. **(c)** Experimental transmission spectra of the near-confocal cavity used for strong coupling of $I_2$ when empty (green) and of the vapor flow cell containing $I_2$ carried in a 90 sccm He flow (magenta, vertically offset for clarity). The frequency axis is referenced with the center of the $I_2$ absorption band at 18788.4405 $cm^{-1}$ corresponding to 0 MHz.

The optical cavity is assembled from two plano-concave aluminum mirrors held concentrically by lens mounts secured to the base of the fiber bench. We use a near-confocal geometry with cavity length $L \sim 1.18$ cm. Near-confocal cavities are stable, easy to align, and feature highly-degenerate Gaussian transverse spatial modes which can be brought simultaneously into resonance with the molecular transition of interest. We fabricate the cavity mirrors by depositing 31 nm of Al on UV-grade fused silica substrates with 1.19 cm radii of curvature (ROC). One cavity mirror is glued to a ring-shaped piezoelectric chip before mounting to enable remote modulation of the cavity length. The cell is fitted with transparent windows to allow optical access along the cavity axis. Note that we must account for ~7 cm of extracavity pathlength through the $I_2$ sample when interpreting our cavity transmission spectra.

Our laser table layout is shown schematically in Figure 2b. Here, we target cavity-coupling near 532.2 nm. Because of the "green gap" of tunable, narrowband cw lasers in this spectral region,[58] we generate green light via second harmonic generation (SHG) of a 1064 nm distributed-feedback diode laser in a PPMgO:LN crystal. An in-fiber optical isolator placed between the laser and the fiber output coupler minimizes optical feedback. We sweep the frequency of the 1064 nm laser via the current driver modulation port with a triangle wave from a function generator to cover >24 GHz (>0.8 $cm^{-1}$) in the green wavelength range. We use most of the green SHG light to perform transmission spectroscopy of the strongly-coupled optical cavity (beam 1 in Figure 2b), spatially mode-matching the green SHG beam to the $TEM_{00}$ cavity mode with a pair of plano-convex lenses.[51] We calibrate the frequency axis of the SHG light with both absolute and relative measures. A small fraction of green light is picked off to monitor transmission of a sealed $I_2$ reference cell for absolute calibration (beam 2 in Figure 2b). For relative frequency calibration, we send residual 1064 nm light after the SHG crystal through a free-space etalon assembled from plano-concave Au mirrors with *ROC* = −21.69 cm placed ~21 cm apart (beam 3 in Figure 2b). More detail on frequency calibration is provided in Section S1.2 of the SI.

We characterize the optical cavity by recording its transmission spectrum when empty (green trace, Figure 2c). Fitting the empty cavity transmission spectrum with eq 1 yields a mirror reflectivity of $R \sim 78.4\%$ near the target wavelength of $\lambda = 532.2$ nm (18788.4 $cm^{-1}$), producing a cavity mode linewidth of $\Delta\nu = 982$ MHz (0.0328 $cm^{-1}$) fwhm and finesse $\mathcal{F} = \pi\sqrt{R}/(1-R) \sim 12.9$. The longitudinal mode order of cavity modes in this spectral region is on the order of $q = 2L/\lambda \sim 44{,}700$. Our cavity mode spacing of ~6330 MHz corresponds to a near-confocal cavity length of 1.18 cm and is half the *FSR* expected for a non-confocal cavity of the same length.[51]

We actively stabilize the cavity length and thereby the frequencies of the cavity modes using a side-of-line locking scheme. We use the transmission of a narrowband metrology-grade cw laser near 1550 nm to generate an error signal, which is then processed with a proportional-integrator loop filter and fed onto a piezoelectric chip which actuates the cavity length (Figure 2b). By slightly tuning the wavelength of the 1550 nm laser, we can walk the cavity length to any desired lock point. This laser locking scheme is described in more detail in Section S1.3 of the SI. Without locking, the cavity is sufficiently stable that a cavity mode stays within one fwhm linewidth of a resonantly-coupled molecular transition for tens of seconds to minutes. However, engaging the lock does prove extremely helpful to reduce noise and minimize averaging over different effective cavity lengths while recording cavity transmission spectra.

To characterize molecular conditions in the cavity cell without removing the cavity mirrors, we replace the cavity cell in beam path 1 with a secondary home-built vapor flow cell of comparable length and attach it to the same inlet and outlet gas lines. We record the optical transmission of the vapor flow cell to understand the number density and pressure broadening of $I_2$ as we vary the He flow rate through the $I_2$ pick-up bottle. The experimental transmission spectrum of this vapor flow cell is shown in magenta in Figure 2c for a He flow rate of 90 sccm.

Some optical feedback in the 1064 nm laser system causes semi-periodic noise in our spectra and can affect lineshapes in both the near-infrared and visible. Oscilloscope averaging of traces over four subsequent frequency sweeps is employed throughout our experiments to help minimize apparent optical feedback. We additionally smooth our spectroscopic traces using a 20-point moving average method, corresponding to a 42 MHz (0.0014 $cm^{-1}$) smoothing window. Figure S5 shows that smoothing preserves peak positions and relative intensities in cavity transmission spectra. This smoothing procedure, in combination with the accuracy of our frequency calibration (see Section S1.2 of the SI) leads to ~50 MHz (0.0017 $cm^{-1}$) uncertainty on all peak positions and Rabi frequencies quoted herein. Finally, all experimental cavity transmission spectra are background corrected and normalized.

## 3. RESULTS AND DISCUSSION

We now discuss cavity coupling of rovibronic transitions in the B–X band of iodine. We target this band because its absorption cross-section is strong and nicely resolved at room temperature, which we previously predicted would enable gas-phase ESC.[54] To aid in quantitative assignment of our data, we construct a rovibronic model of the $I_2$ B–X band in the PGOPHER software package.[59] This model is informed by a PGOPHER example model of the $I_2$ hyperfine structure[60] and incorporates the best available vibrational frequencies, rotational constants, hyperfine parameters, dipole moments, and Franck-Condon factors from the literature.[61–67] We ensure agreement of the model with available broadband, low resolution[57] and narrowband, high-resolution[61] data, only making small adjustments to band strengths to match experimental signatures. More details on the $I_2$ model are provided in Section S2.1 of the SI.

The $I_2$ B–X band consists of an extended Franck-Condon vibronic progression (Figure 3a), which arises because the $I_2$ bond length is much longer in the B excited electronic state than the X ground electronic state. Contributions from transitions that begin in the $\nu_1'' = 0$ vibrational ground state of the X state are shown in cyan in Figure 3a. The higher-resolution spectra in Figures 3b and 3c illustrate the individual *ro*vibronic transitions that underlie the peaks in the Franck-Condon progression. The spectrum is quite congested with significant overlap between clusters of transitions belonging to single rovibronic peaks in the progression, as evidenced by the transition labels in Figure 3c.

We target cavity-coupling of a peak at 18788.4405 $cm^{-1}$ (532.242152 nm) indicated with magenta arrows in Figure 3, which arises from two overlapping B–X rovibronic transitions: $\nu_1'' = 0 \rightarrow \nu_1' = 32$, $J'' = 53 \rightarrow J' = 52$ and $\nu_1'' = 0 \rightarrow \nu_1' = 34$, $J'' = 103 \rightarrow J' = 102$, where $\nu_1''$ and $J''$ are the vibrational and rotational quantum numbers of the X electronic ground state and $\nu_1'$ and $J'$ are those of the B electronic excited state. These transitions are hereafter referred to as 32-0 P(53) and 34-0 P(103). The frequency axes of all cavity transmission spectra reported herein are referenced with the center of this target 18788.4405 $cm^{-1}$ transition corresponding to 0 MHz.

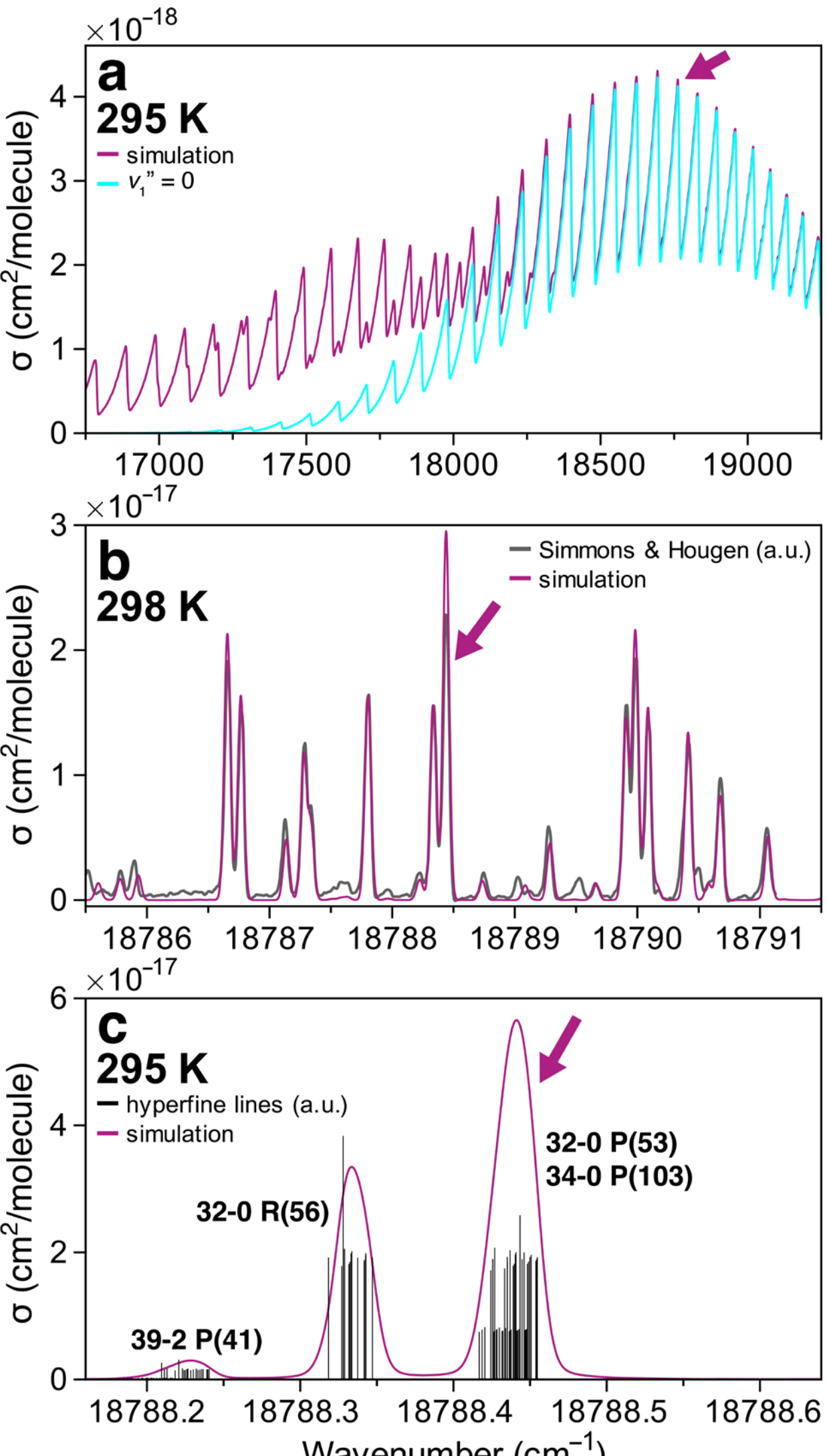

**Figure 3. (a)** Absorption cross section of the $I_2$ B–X band at 295 K. The rovibronic model for $I_2$ is built in PGOPHER (magenta) and fit to experimental air-broadened data from Saiz-Lopez *et al.*[57] with a Gaussian linewidth of 5 $cm^{-1}$ fwhm and a Lorentzian linewidth of 0.4 $cm^{-1}$ fwhm. Contributions from transitions originating in the $\nu_1'' = 0$ vibrational ground state of the electronic ground state are shown in cyan. **(b)** The PGOPHER $I_2$ model broadened with a Gaussian linewidth of 0.055 $cm^{-1}$ fwhm and a Lorentzian linewidth of $1 \times 10^{-4}$ $cm^{-1}$ fwhm (magenta) reproduces rovibronic line positions of the experimental absorption data of Simmons and Hougen[61] (gray). We have digitized and smoothed the Simmons and Hougen data which reported an instrumental resolution of 0.055 $cm^{-1}$; the original dataset is reported with arbitrary intensity units and strong transitions may be saturated. **(c)** The hyperfine structure (black) of the $I_2$ transitions of interest is unresolved under the expected experimental broadening conditions (magenta). Here, we simulate the spectrum with 0.0145 $cm^{-1}$ fwhm Gaussian Doppler broadening and 0.00319 $cm^{-1}$ fwhm Lorentzian pressure broadening. Rovibronic transitions are labeled with vibrational and rotational quantum numbers. The $I_2$ transition of interest for strong coupling in the present work is marked with a magenta arrow in each panel.

A representative extracavity transmission spectrum of the $I_2$ vapor flow cell is shown in Figure 4a, featuring B−X rovibronic transitions in the spectral region of interest. The lineshapes of the individual molecular transitions are dominated by unresolved hyperfine states (see Figure 3c), yielding a linewidth of 988 MHz (0.0329 cm$^{-1}$) fwhm. We provide more details on simulating extracavity transmission spectra and experimental broadening conditions in Section S2.2 of the SI. In brief, the molecular lineshape is well captured by convolving the hyperfine lines with a 435 MHz (0.0145 cm$^{-1}$) fwhm Gaussian lineshape to account for room-temperature Doppler broadening and a Lorentzian lineshape from pressure broadening whose linewidth depends on the He flow rate. We observe a maximum Lorentzian broadening of 215 MHz (0.00716 cm$^{-1}$) fwhm from air- and He-broadening of the $I_2$ lines at the highest cavity cell flow rate of 100 sccm; $I_2$ self-broadening is expected to be comparatively small. We ignore transit time broadening here which we expect to be minimal, contributing at most ~2.2 MHz fwhm Lorentzian broadening, assuming a 32 μm cavity waist (1/e$^2$ radius).

To achieve strong coupling, the Rabi splitting must exceed both the molecular and cavity mode linewidths. Here, we take the molecular linewidth to be a reasonable target for the design cavity linewidth. As described above in the Experimental Section, our near-confocal cavity features a mode linewidth of $\Delta\nu$ = 982 MHz fwhm, a close match for the 988 MHz fwhm linewidth of the $I_2$ peak of interest. Note that we achieve a narrow cavity linewidth here using mirrors with modest reflectivity, $R \sim 78.4\%$. The standard Fabry-Pérot cavity linewidth is given by $\Delta\nu = FSR\ /\ \mathcal{F}$, where $\mathcal{F}$ is set only by the mirror reflectivity. For fixed $\mathcal{F}$, a longer cavity features a smaller $FSR$ and therefore a smaller $\Delta\nu$. GHz-scale cavity linewidths are therefore easily accessible in the long, centimeter-scale cavity used here, but would require much higher-finesse mirrors in a microcavity geometry.

With well-matched cavity and molecular linewidths to work with, we tune the cavity length to bring a photonic mode into resonance with the $I_2$ transition of interest at 18788.4405 cm$^{-1}$. By increasing the flow rate of He carrier gas and therefore the intracavity [$I_2$], we observe the attenuation then splitting of the cavity mode (Figure 4b). Cavity mode splitting begins to be resolvable with a 12 sccm He flow rate; with a 16 sccm He flow, the mode splitting of 1103 MHz exceeds both the cavity and molecular mode linewidths. We observe a maximum Rabi splitting of 1722 MHz at a 100 sccm He flow rate, comfortably within the strong coupling regime. We confirm that the observed mode splitting is independent of the incident laser power used to monitor the cavity transmission (see Figure S6).

We use the classical Fabry-Pérot cavity transmission equation (eq 1) to simulate the experimental spectra and extract the intracavity $I_2$ number density for each trace; these simulated spectra are plotted in Figure 4c, color-coded to correspond to the experimental spectra. To fit each experimental spectrum, we use the rovibronic $I_2$ model to simulate the absorption cross-section $\sigma(\nu)$ specific to the pressure broadening conditions at the relevant flow rate and float $R$, $L$, and $N/V$ in eq 1. We assume a homogeneous $I_2$ concentration over the whole cavity volume and account for free-space absorption by $I_2$ in the ~7 cm of extracavity pathlength in the cell through which the cw light still passes. The fits return sensible values for $R$ and $L$, consistent with the known experimental parameters laid out in Section 2 above. We do observe a drop in $R$ from 78.4% to 76.3% over the course of flow experiments as the cavity cell is filled with higher number densities of $I_2$. We believe that this reduced mirror performance is a result of the adsorption[68] of $I^-$ and $I_2$ by the thin native $Al_2O_3$ oxide layer that forms on Al surfaces after exposure to air.[69,70] Fit values of $N/V$ are also reasonable based on comparison to extracavity $I_2$ measurements in the vapor flow cell, and, as expected, [$I_2$] plateaus near the room-temperature vapor pressure at high flow rates (see legend of Figure 4c).

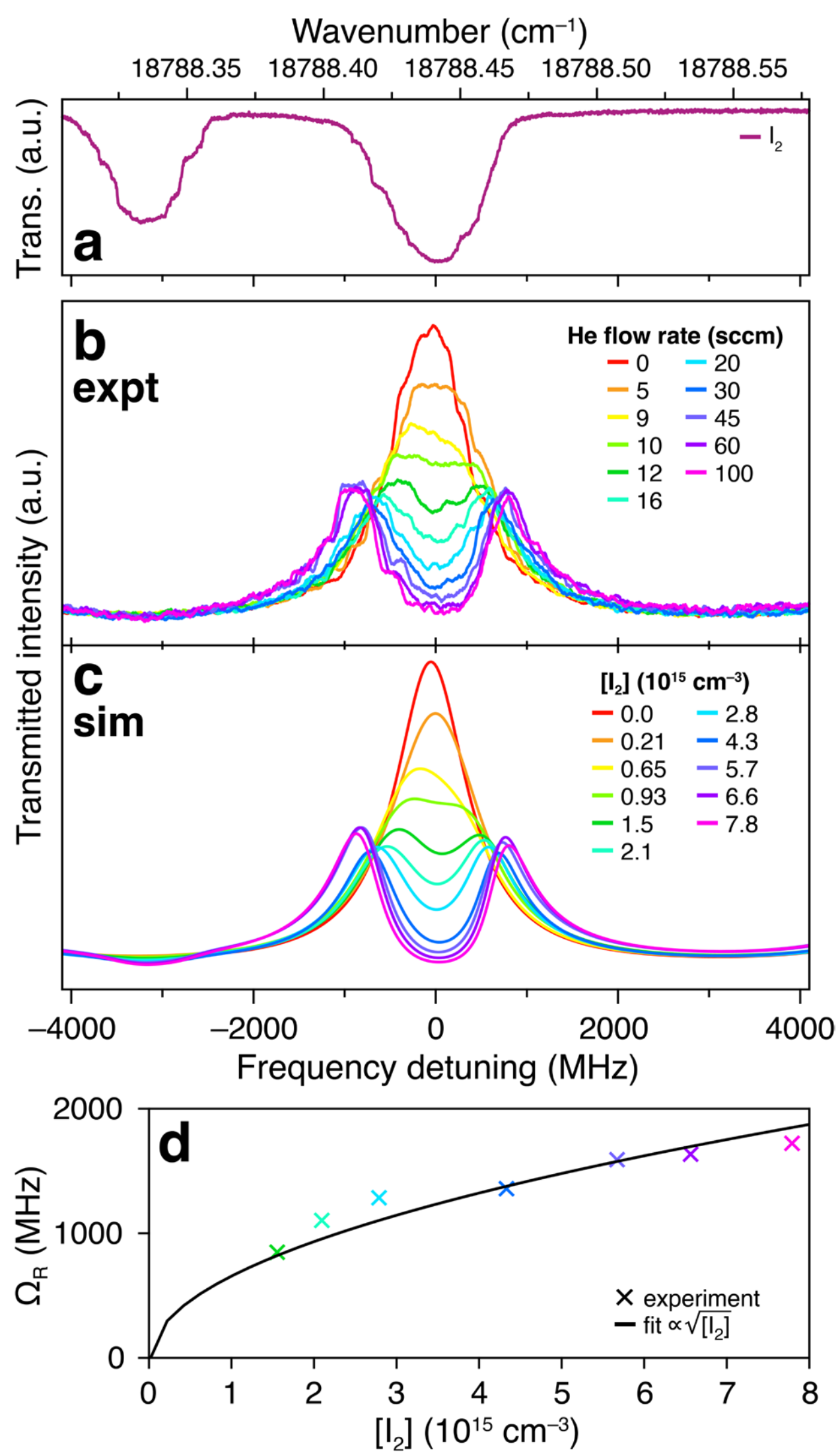

**Figure 4. (a)** Transmission spectrum of extracavity vapor flow cell containing $I_2$ carried in a 90 sccm flow of He. The frequency axis is referenced with the center of the absorption band at 18788.4405 cm$^{-1}$ corresponding to 0 MHz. **(b)** Experimental transmission spectra of the near-confocal Fabry-Pérot cavity under systematic tuning of the intracavity $I_2$ number density via the He carrier gas flow rate. **(c)** Simulated cavity transmission spectra obtained via fitting of the corresponding experimental traces in panel (b) with the Fabry-Pérot cavity transmission expression in eq 1. **(d)** Rabi splittings extracted from the experimental spectra are plotted as a function of [$I_2$] with color-coding to match panels (b) and (c); data that meet the strong coupling condition are fit with a function proportional to $[I_2]^{1/2}$ (black).

We plot the mode splitting extracted from each experimental trace as a function of [$I_2$] in Figure 4d, highlighting the $\hbar\Omega_R \propto [I_2]^{1/2}$ dependence characteristic of the collective strong light-matter coupling regime. The maximum Rabi splitting of 1722 MHz achieved with 100 sccm of He carrier gas corresponds to [$I_2$] ~ 7.8 × $10^{15}$ $cm^{-3}$, within experimental uncertainty of the room-temperature $I_2$ vapor pressure. We could reach higher [$I_2$] and consequently larger $\hbar\Omega_R$ values by heating the $I_2$ sample, gas lines, and gas cell, but we are limited to room-temperature experiments in our current configuration.

We next consider how the cavity transmission spectrum behaves as the cavity mode is systematically detuned from resonance with the $I_2$ peak at 18788.4405 $cm^{-1}$. In planar microcavities that support a continuum of transverse modes with finite in-plane momentum, dispersion plots are usually generated by tuning the incident angle of radiation. Here, it is easiest to tune the discrete modes supported by our near-confocal cavity by systematically stepping the cavity length via piezoelectric actuation, maintaining on-axis laser coupling into the $TEM_{00}$ mode throughout. We plot experimental detuning spectra acquired at a constant 30 sccm He flow rate corresponding to [$I_2$] ~ 4.3 × $10^{15}$ $cm^{-3}$ in Figure 5a, alongside simulations in Figure 5b. The spectra shown in Figure 5a are also plotted individually in Figure S2. A characteristic avoided crossing of the polariton modes is evident as the cavity mode passes through resonance with the molecular transition at 0 MHz detuning. We also observe attenuation of the cavity mode near resonance with the $I_2$ 32-0 R(56) transition centered at 18788.3340 $cm^{-1}$ (–3193 MHz detuning). In a separate set of experiments shown in Figure S7, we directly target cavity-coupling of the 32-0 R(56) rovibronic transition but do not quite reach the strong coupling regime given the weaker absorption cross-section of this transition and the limited [$I_2$] accessible in the current set-up.

In summary, we have demonstrated strong light-matter coupling between a Fabry-Pérot cavity mode and rovibronic transitions of gas-phase molecular iodine. We actively maintain resonance conditions throughout flow experiments by stabilizing the cavity length, a level of control also shown in a recent nanofluidic cavity experiment[33] but uncommon in the solid-state cavities usually harnessed for ESC. We achieve reasonable collective coupling strengths by working with the room-temperature vapor pressure of $I_2$ and expect higher number densities to be accessible in future work by heating solid $I_2$ to increase its vapor pressure. Other rovibronic transitions of $I_2$ or other gas-phase molecules with strong absorption cross-sections (>4×$10^{-17}$ $cm^2$/molecule) and narrow linewidths should be similarly accessible for gas-phase ESC. Next-generation cavity designs with much smaller mode volumes, like fiber-tip cavities[71] or photonic crystals,[72] could also be used to increase the cavity-coupling strength per molecule.

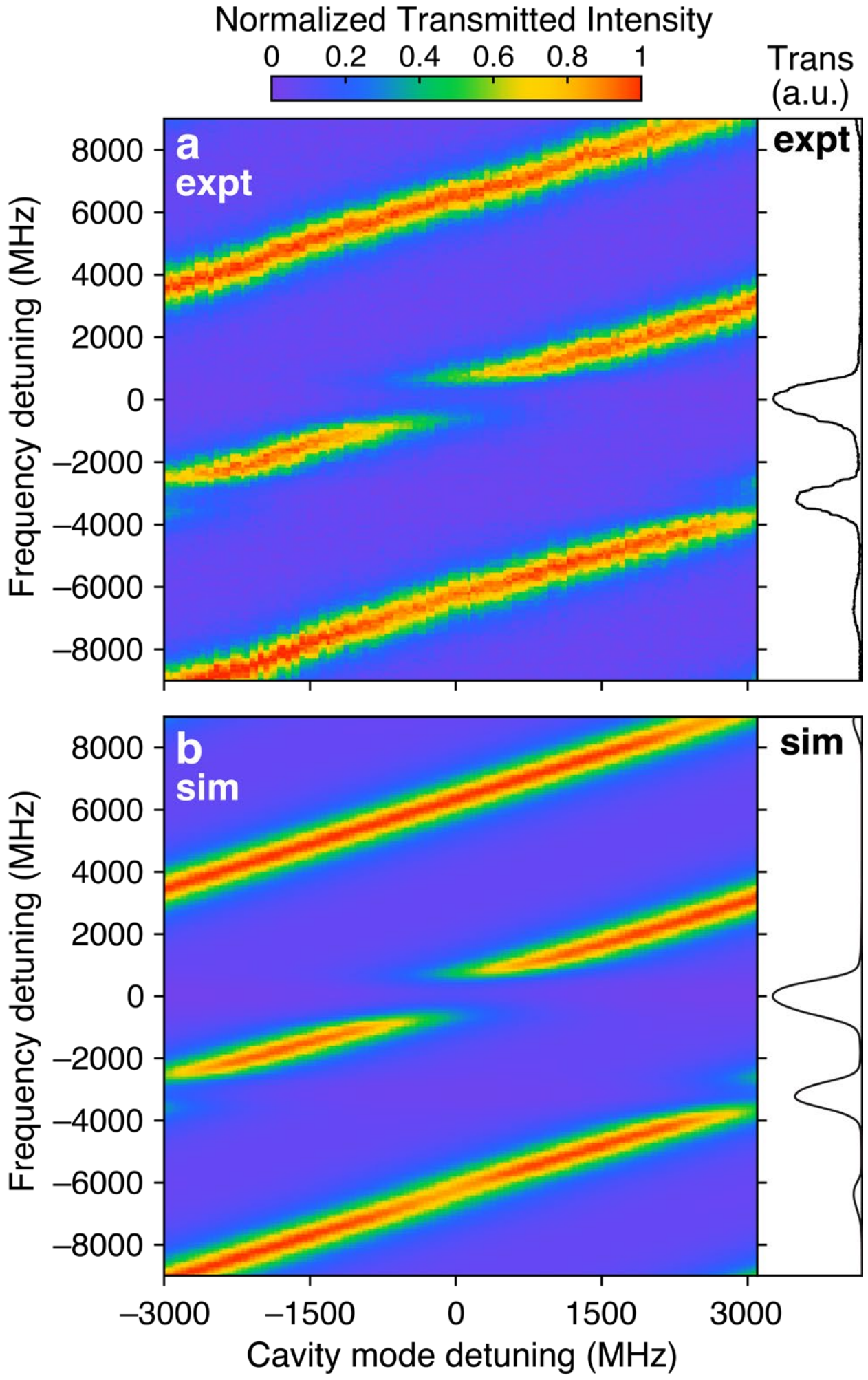

**Figure 5. (a)** Dispersion of the experimental transmission spectra for the near-confocal Fabry-Pérot cavity containing $I_2$ carried in 30 sccm of He. The cavity length is systematically stepped so a photonic mode is brought through resonance with the $I_2$ transition of interest at 18788.4405 $cm^{-1}$. Both frequency axes are referenced with the center of this $I_2$ absorption band corresponding to 0 MHz. The experimental extracavity transmission spectrum of $I_2$ in a 90 sccm He flow is plotted in black for reference. **(b)** Simulated dispersion plot of the cavity transmission spectra obtained using parameters from experimental fitting and eq 1. The simulated transmission spectrum of [$I_2$] = 3.2 × $10^{15}$ $cm^{-3}$ over 8 cm of path length is plotted for reference in black.

While the current work represents a proof-of-principle demonstration, we expect gas-phase molecular ESC to open new opportunities in polariton photophysics and photochemistry. We reach the strong coupling regime in an ensemble of isolated molecules that will not form aggregates or excimers as seen in the dense films more commonly used for ESC.[36] Our centimeter-scale cavity geometry will also permit optical access orthogonal to the cavity axis, allowing for modulation and interrogation of intracavity species without the challenges of optical interference inherent in on-axis spectroscopy. Examining the nonlinear spectroscopy, fluorescence, and photodissociation of $I_2$ as a function of cavity-coupling conditions should be highly feasible in future work. Controlled studies of po-lariton lasing and condensation may also be accessible with this platform. Observations of such benchmark processes may ultimately aid in advancing mechanistic understanding of molecular dynamics under ESC.

## ASSOCIATED CONTENT

### Supporting Information

Experimental methods, simulation and data processing details, and additional experimental cavity transmission spectra and simulations (PDF)

## AUTHOR INFORMATION

### Corresponding Author

* **Marissa L. Weichman** – *Department of Chemistry, Princeton University, Princeton, New Jersey 08544, United States*; orcid.org/0000-0002-2551-9146; email: weichman@princeton.edu

### Authors

**Jane C. Nelson** – *Department of Chemistry, Princeton University, Princeton, New Jersey 08544, United States;* https://orcid.org/0000-0002-2560-2775

**Trevor H. Wright** – *Department of Chemistry, Princeton University, Princeton, New Jersey 08544, United States;* https://orcid.org/0000-0002-9330-8631

**Neo Lin** – *Department of Chemistry, Princeton University, Princeton, New Jersey 08544, United States;* https://orcid.org/0000-0002-0535-564X

**Madeline Rohde** – *Department of Chemistry, Princeton University, Princeton, New Jersey 08544, United States;* https://orcid.org/0009-0004-0192-5862

### Funding

This research was supported by the Air Force Office of Scientific Research (AFOSR) under grant FA9550-25-1-0157. J.C.N. acknowledges support from a National Defense Science and Engineering Graduate research fellowship. M.R. acknowledges support from the Leach Summer Scholars Program.

### Notes

The authors declare no competing financial interest.

## ACKNOWLEDGMENT

The authors acknowledge the use of the Micro/Nano Fabrication Center (MNFC), a core shared-use facility of the Princeton Materials Institute (PMI), for cavity mirror fabrication. The authors acknowledge helpful conversations with David Chandler.

# Supporting Information:

# Electronic Strong Coupling of Gas-Phase Molecular Iodine

*Jane C. Nelson, Trevor H. Wright, Neo Lin, Madeline Rohde, and Marissa L. Weichman**

Department of Chemistry, Princeton University, Princeton, New Jersey 08544, United States

* weichman@princeton.edu

# S1. Experimental methods

## S1.1. Optical cavity cell and vapor flow cell

Our home-built cavity cell is illustrated in Figure 2a of the main text. Inspired by previous implementations in the literature,[1,2] the cavity is mounted on a fiber bench (Thorlabs FB-76W) whose cover is glued in place with epoxy (Torr Seal) to form a sealed vapor cell. ¼” steel Swagelok connections are glued to holes drilled in the top of the cover to provide a gas inlet and outlet. Because of the inevitable exposure to corrosive $I_2$ gas, we evacuate the cavity cell with an inexpensive oil roughing pump (SPECSTAR) and reach a modest base pressure of ~14 torr in the cell. For iodine experiments, we deliver $I_2$ to the cavity cell by passing helium carrier gas (Airgas) through a sealed pick-up cell of $I_2$ crystals (Sigma-Aldrich) via ¼” tubing. The He flow is varied from 0 to 100 sccm with a mass flow controller (Alicat Scientific). We use a pressure gauge (Kurt J. Lesker) to characterize the cavity cell pressure for different He flow rates, though this gauge is not compatible with iodine and can therefore not be used during strong coupling experiments.

Optical access to the cavity cell is made possible via anti-reflection coated $CaF_2$ flats (Edmund Optics 46-098) held with o-rings in lens mounts (Thorlabs SM1L03) that are glued to the end plates of the fiber bench. The cavity mirrors themselves are mounted on the fiber bench in lens tubes (Thorlabs SM5L05) aligned concentrically in lens tube mounts (Thorlabs FBS05) that are secured to the fiber bench base. One mirror is glued to a piezoelectric chip (Thorlabs PA44M3KW) which is in turn glued to its lens tube. Each cavity mirror consists of 31 nm of aluminum deposited on a 1.19 cm radius of curvature plano-concave UV-grade fused silica substrate. We use an Angstrom Nexdep E-beam Evaporator located in the Princeton Micro/Nanofabrication Center (MNFC) for aluminum deposition.

Upon exposure to $I_2$, we observe loss of reflectivity in the cavity mirrors from $R = 78.4\%$ to 76.3%, corresponding to a change in cavity mode linewidth from $\Delta\nu = 982$ to 1099 MHz. The Al mirror coatings likely develop a native $Al_2O_3$ surface oxide layer a few nanometers thick upon exposure to air.[3,4] This $Al_2O_3$ surface can then adsorb $I^-$ and $I_2$,[5] especially at high intracavity $I_2$ number densities approaching the $7.4 \times 10^{15}$ $cm^{-3}$ room-temperature vapor pressure,[6] leading to increased absorption of 532 nm light by the mirrors. We see the drop in mirror reflectivity and overall transmission within seconds upon introduction of higher He flow rates and $I_2$ densities to the cell; it then takes a few hours of evacuating the chamber for the system to return to its original state.

To characterize extracavity molecular conditions — as it is difficult to remove the mirrors from the cavity cell — we use a second cell of comparable length to our cavity fiber bench assembly, referred to throughout the text as the vapor flow cell. The vapor flow cell is constructed from a ~7.5 cm long aluminum block with a ~2.6 cm diameter central bore. The cell features similar gas-line connections and optical access to the cavity cell and can therefore be easily swapped into the experiment. When evacuated with the roughing pump, the vapor flow cell reaches a comparable base pressure to the cavity cell.

## S1.2. Laser spectroscopy and frequency calibration

Our laser table layout is shown in Figure 2b of the main text. We use a tunable continuous-wave (cw) distributed feedback near-infrared diode laser centered near 1064 nm (QPhotonics). This laser features a narrowband < 3 MHz instantaneous linewidth that can be tuned from 1061.6 to 1064.7 nm using current and temperature controllers (Newport 505B, Newport 8610.8C). We collect spectra by sweeping the laser frequency with a 10 Hz triangle wave from a function generator (BK Precision) to the current driver modulation port. The 1064 nm laser is fiber-coupled into an isolator (Thorlabs IO-G-1064-APC) then coupled into free space with an output coupler (Thorlabs F230APC-1064).

We generate green 532 nm light for cavity transmission spectroscopy via second harmonic generation (SHG) of the 1064 nm light. Free-space 1064 nm light is focused into a PPMgO:LN bulk chip SHG crystal (HC Photonics SHVIS-SB-50). The crystal is temperature-controlled in an oven (HC Photonics TC-038D) to match the quasi-phase matching period to the wavelength of the incoming 1064 nm beam. We optimize polarization, steering, and crystal temperature to achieve ~40 μW of 532 nm power from ~30 mW of 1064 nm light. After the crystal, residual 1064 nm light is separated from the SHG beam using a shortpass dichroic mirror (Thorlabs DMSP900). A second shortpass dichroic mirror placed in the transmitted 532 nm beam further suppresses residual 1064 nm, isolating narrowband green light.

Most of the 532 nm light generated from SHG (~82%) is directed to the cavity cell (beam 1 in Figure 2b) using a UV-grade fused silica flat coated with 31 nm Al as a beamsplitter (homemade during the same coating run in which we make Al cavity mirrors). We use plano-convex lenses to mode match the 532 nm beam into the $TEM_{00}$ spatial mode of the optical cavity. We read out the cavity transmission with a Si variable-gain avalanche detector (Thorlabs APD430A2).

A small portion (~3%) of the 532 nm light is directed to a reference iodine vapor cell (Thorlabs GC19100-I) for absolute frequency calibration (beam 2 in Figure 2b). We detect the transmission of green light through the vapor cell with a Si detector (Thorlabs DET36A2). This Si detector has a slow rise time, which we account for by shifting the reference vapor cell time-domain signal by 250 μs. We calibrate this correction by synchronously acquiring spectra in the reference cell in beam 2 and the vapor flow cell in beam 1 and ensuring that the $I_2$ absorption peaks coincide in time.

The residual 1064 nm light is collimated with a lens and directed to a free-space etalon for relative frequency calibration (beam 3 in Figure 2b). This etalon is ~21 cm long and assembled from two homemade gold-coated mirrors with reflectivity $R \sim 87\%$. Each mirror consists of an 11 nm layer of Au deposited on a 21.69 cm radius of curvature plano-concave $CaF_2$ substrate using the Angstrom Nexdep E-beam Evaporator in the Princeton MNFC. Light transmitted through the etalon is collected with a Ge detector (Thorlabs PDA30B2). The near-confocal Au etalon features an *FSR* of around 360 MHz, which provides sufficiently fine frequency calibration for the present measurements.

To generate a calibrated frequency axis, we first use the laser sweep parameters to convert time to laser current. Figure S1a shows the raw Au etalon trace (beam 3) as a function of laser current. We process the etalon spectrum by background subtraction of a linear slope, and bandpass filtering to correct for optical feedback in the 1064 nm beamline (Figure S1b). We then extract the positions of extrema in the processed etalon data to map laser current to a relative frequency axis based on an initial guess for the etalon *FSR*.

We next use the simultaneously-collected $I_2$ reference vapor cell transmission spectrum (beam 2) as an absolute frequency reference to optimize the initial guess for the etalon *FSR*. The laser sweep range covers two major $I_2$ absorption peaks known to lie at 18788.3340 and 18788.4405 $cm^{-1}$ (Figure S1c). The positions of these features allow us to offset our experimental frequency axis by an absolute frequency. In addition, we evaluate if the relative frequency axis set by the etalon *FSR* over or underestimates the spacing between the $I_2$ peaks (0.1065 $cm^{-1}$, 3193 MHz) and adjust our initial guess for the etalon *FSR* to correct for the mismatch. When the calibrating optimization loop has minimized the error in iodine peak positions to <10 MHz, we consider the frequency calibration complete.

Finally, we generate a reliable intensity scale for our cavity transmission spectra by accounting for the changing laser intensity throughout the frequency sweep. The laser intensity is frequency dependent due to a combination of power-current nonlinearity in the 1064 nm laser itself and the efficiency of SHG at a fixed oven temperature. We fit the background of the $I_2$ reference vapor cell spectrum with a second-order polynomial and normalize all 532 nm transmission spectra accordingly. Further, we scale cavity transmission spectra by a constant such that a selected off-resonance cavity mode features the same intensity across all spectra. This normalization helps account for changing optical losses in the system as $I_2$ adsorbs on various surfaces in the cell.

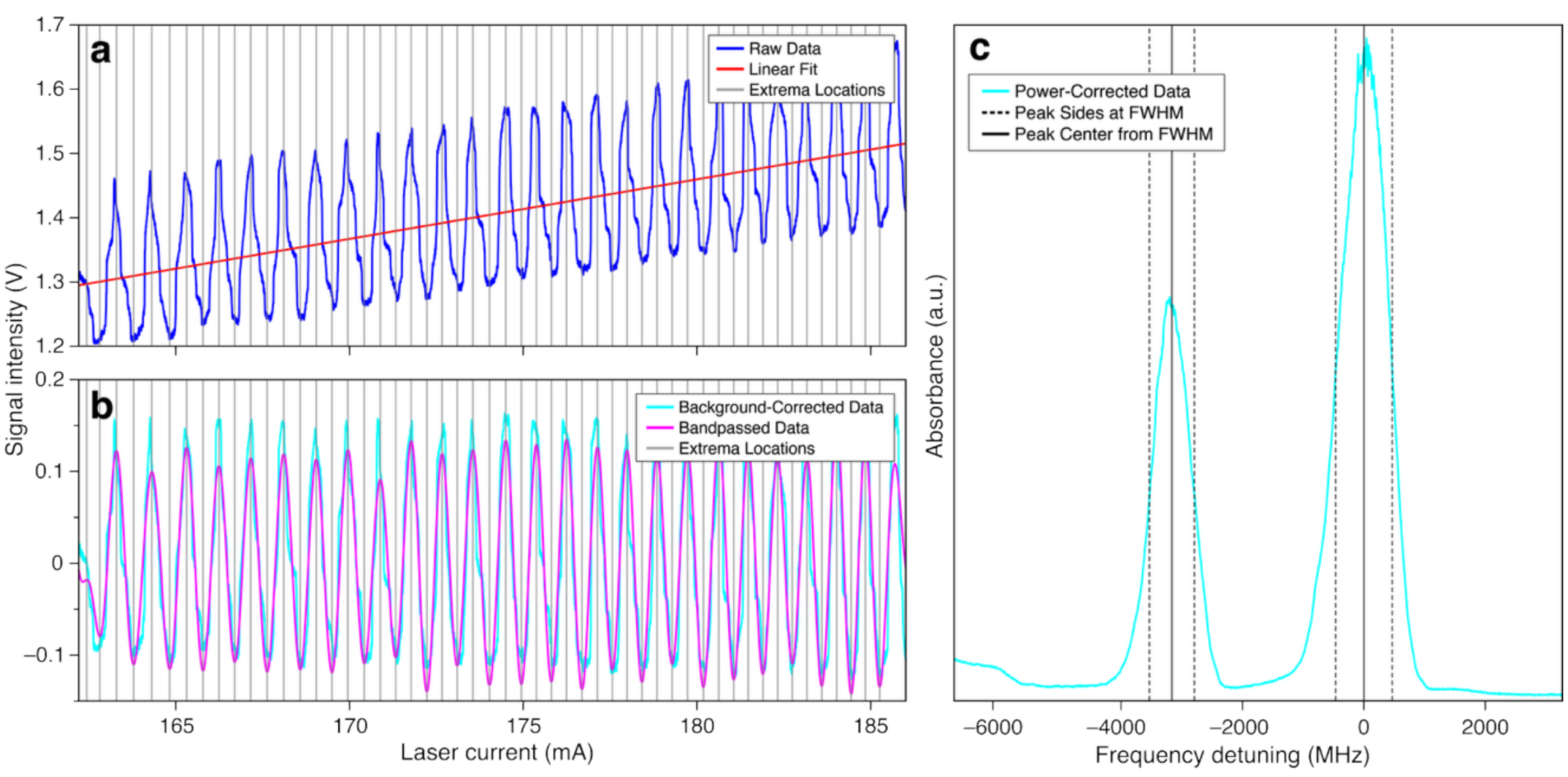

**Figure S1. (a)** Raw 1064 nm transmission though the Au etalon (beam 3) over the laser current sweep (blue). The center of the trace is fit with a linear slope (red). **(b)** We correct the raw etalon data by subtracting the linear background signal (cyan) and applying a bandpass filter to reduce the spectral artifacts from optical feedback (pink). Extracted extrema locations from the corrected data are plotted as gray vertical lines in panels (a) and (b). **(c)** Transmission of 532 nm light through the $I_2$ reference cell is corrected for changing light intensity over the laser sweep and converted to absorbance. We find the absorption peak centers by finding the mid-point of the full-width half maxima (fwhm), which is relatively insensitive to the asymmetric envelope of hyperfine lines and lineshape distortion from optical feedback. We generate an absolute frequency calibration by ensuring that these peak centers are consistent with those in our rovibronic model of $I_2$. The frequency axis is referenced with the center of the $I_2$ absorption band at 18788.4405 $cm^{-1}$ corresponding to 0 MHz.

## S1.3. Cavity length stabilization and tunability

We employ a side-of-line locking scheme to stabilize the cavity length to fixed setpoints, reducing the effects of vibrational noise and long-term drift. The optical locking layout is shown in Figure 2b of the main text. One cavity mirror is glued to a piezoelectric chip to allow small corrections in cavity length. We generate the error signal for the cavity lock using a tunable metrology-grade cw laser near 1550 nm (RIO ORION, 4 kHz linewidth) as a frequency reference far from resonance with any $I_2$ lines. The RIO output is aligned through the optical cavity, concentric with the 532 nm spectroscopy beam. The two beams are overlapped before and separated after the cavity with longpass dichroic mirrors (Thorlabs DMLP950). The intensity of 1550 nm light transmitted through the cavity is detected with an amplified Ge detector (Thorlabs PDA30B2) and processed with a proportional-integral loop filter (New Focus LB1005) to generate an error signal which is then fed to the cavity piezo via its controller (Thorlabs MDT694B). We can readily lock the system so the side of a cavity fringe remains resonant with the RIO laser, as the laser linewidth is orders of magnitude narrower than the cavity linewidth. We can also step the cavity length lock point by stepping the RIO laser wavelength via the thermistor resistance of its thermoelectric cooler and thereby move cavity fringes in and out of resonance with $I_2$ transitions near 532 nm. When the RIO wavelength is stepped sufficiently slowly, the cavity remains locked and follows the laser to stabilize at each new cavity length.

We collect cavity transmission spectra at systematically stepped cavity lengths to construct dispersion plots, as shown in Figure 5a of the main text. The individual spectral traces underpinning Figure 5a are plotted separately below in Figure S2. A dispersion plot is also provided for an empty cavity containing no $I_2$ in Figure S3.

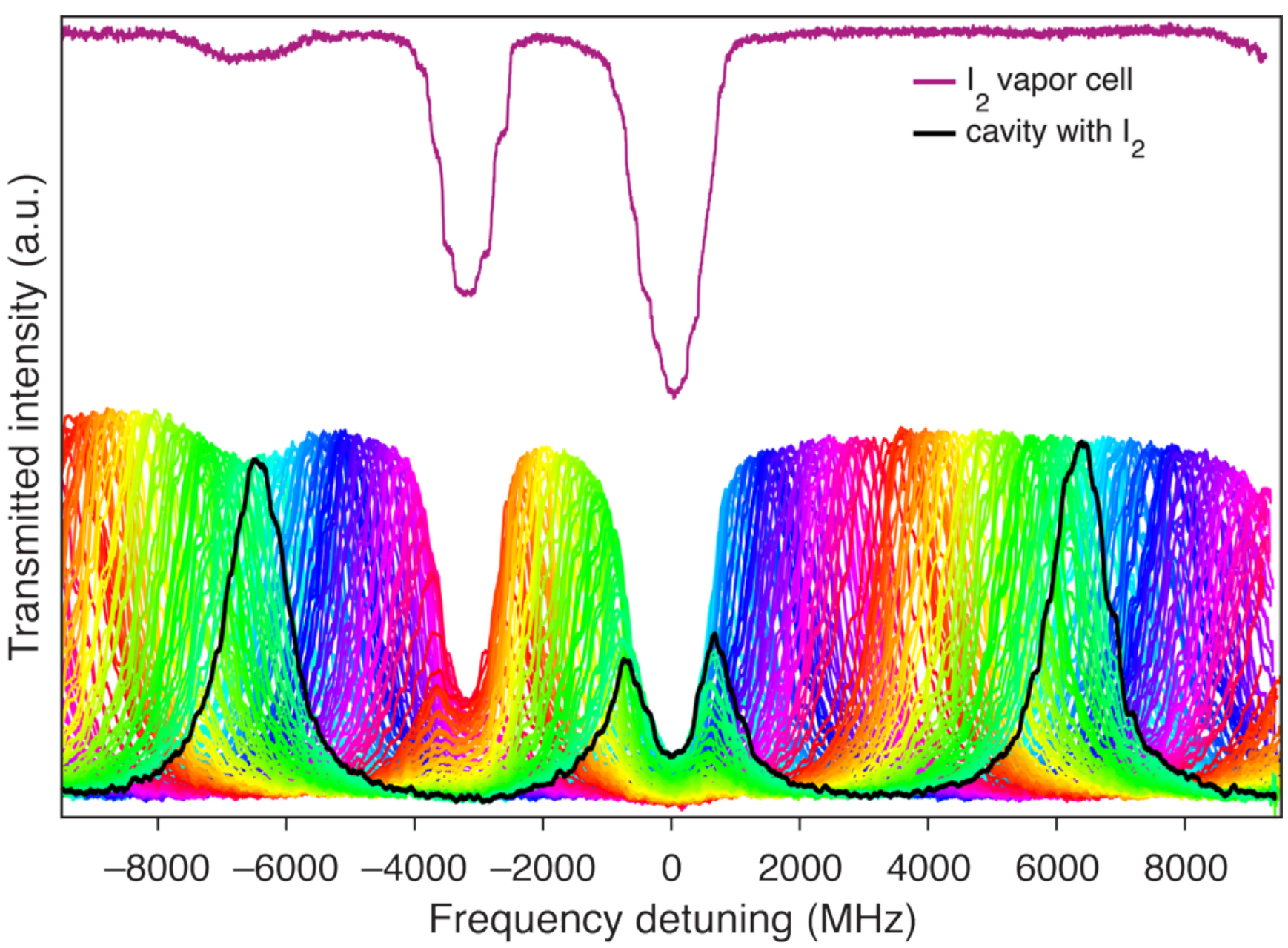

**Figure S2.** Experimental transmission spectra (colored traces) of the near-confocal Fabry-Pérot cavity containing $I_2$ in a 30 sccm flow of He that make up the dispersion plot shown in Figure 5a of the main text. The cavity length is systematically stepped so a photonic mode is brought through resonance with the $I_2$ transition of interest; the near-resonance spectrum is plotted in black. The frequency axis is referenced with the center of the $I_2$ absorption band at 18788.4405 cm$^{-1}$ corresponding to 0 MHz. The experimental transmission spectrum of extracavity $I_2$ is plotted in magenta for reference (offset for clarity).

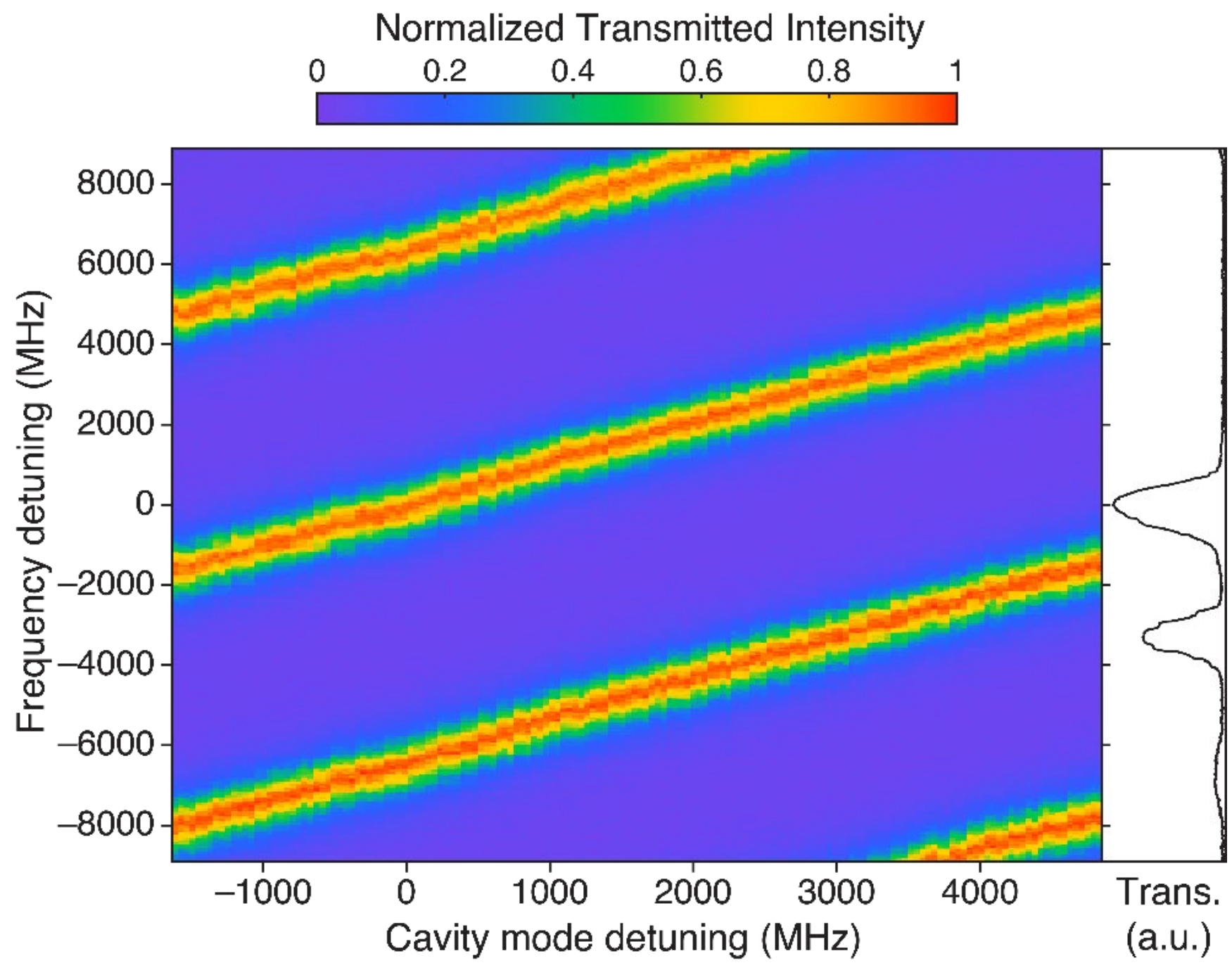

**Figure S3.** Dispersion plot of experimental transmission spectra for the empty near-confocal Fabry-Pérot cavity. The cavity length is systematically stepped so the photonic modes are scanned in frequency. Both frequency axes are referenced with the center of the $I_2$ absorption band at 18788.4405 cm$^{-1}$ corresponding to 0 MHz. The experimental transmission spectrum of extracavity $I_2$ is plotted in black for reference.

# S2. Modeling and data processing

## S2.1. Rovibronic model of iodine absorption cross-section

Simulating cavity transmission spectra using eq 1 of the main text requires as inputs the cavity parameters $R$, $T$, and $L$, the intracavity molecular number density $N/V$, and the molecular absorption cross-section, $\sigma(\nu)$. In our prior work on methane,[7–9] a detailed line list was available from the HITRAN database,[10] making simulations of $\sigma(\nu)$ straightforward with the PGOPHER[11] software package. Unfortunately, no such line list is available for iodine. Instead, we construct a model for the rovibronic structure of the $I_2$ B−X band in PGOPHER based on published experimental spectroscopic constants[12–17] and an existing PGOPHER model for the $I_2$ hyperfine structure.[18] We float less established parameters including lineshape broadening values and band intensities to fit this model to available broadband, low resolution[6] and narrowband, high-resolution[12] reference spectra, as illustrated in Figure 3 of the main text.

To generate the $I_2$ absorption cross-section to simulate a given cavity transmission spectrum, we convolve the PGOPHER model with the relevant lineshape given experimental conditions. As all experiments are performed at room temperature (295 K), we convolve with a Gaussian lineshape with 435 MHz (0.0145 cm$^{-1}$) fwhm to capture Doppler broadening in all cases. We also treat pressure broadening via convolution with a Lorentzian lineshape whose width is determined by the relevant experimental background pressure for a given experiment, as described below in Section S2.3.

## S2.2. Extracavity vapor flow cell pressure broadening and data processing

We swap out the cavity cell with an extracavity vapor flow cell to characterize the intracavity molecular conditions, as it is difficult to determine these conditions with cavity mirrors in place. We vary the flow rate of He carrier gas through the $I_2$ sample to tune the molecular number density in the cell. We record transmission spectra at each flow rate and convert to absorbance before fitting with our PGOPHER model. We find excellent agreement between the experimental and simulated data for the two $I_2$ absorption features at 18788.3340 and 18788.4405 $cm^{-1}$ when appropriate broadening is included in the simulation (Figure S4). For all spectra, we hold the Gaussian broadening at 435 MHz (0.0145 $cm^{-1}$) fwhm corresponding to Doppler broadening at 295 K. We must additionally include Lorentzian broadening which arises primarily from pressure broadening and increases at higher He flow rates.

We use the following $I_2$ pressure broadening coefficients compiled from the range of reported literature values:[19–24] $\gamma_{air}$ = 0.25 $cm^{-1}$ fwhm/atm, $\gamma_{He}$ = 0.10 $cm^{-1}$ fwhm/atm, and $\gamma_{self}$ = 0.32 $cm^{-1}$ fwhm/atm. As we do not have an appropriate iodine-resistant pressure gauge, we calibrate the pressures associated with each He flow rate in the absence of $I_2$ in both the cavity cell and vapor flow cell. We then estimate partial pressures of air and He which we in turn use to calculate the air- and He-broadening expected for $I_2$ at each He flow rate. We ignore self-broadening since even at our highest $I_2$ number densities, nearing $[I_2] = 8 \times 10^{15}$ $cm^{-3}$, the expected contribution of self-broadening ($9.5 \times 10^{-5}$ $cm^{-1}$ or 2.8 MHz) is small compared to other sources of broadening. We confirm that our estimated pressure broadening accurately reflects the vapor flow cell $I_2$ spectra as shown in Figure S4 for a 90 sccm He flow rate. We perform the same pressure calibration in the cavity cell to calculate the $I_2$ absorption cross-section used in simulating the cavity transmission spectrum at each flow rate.

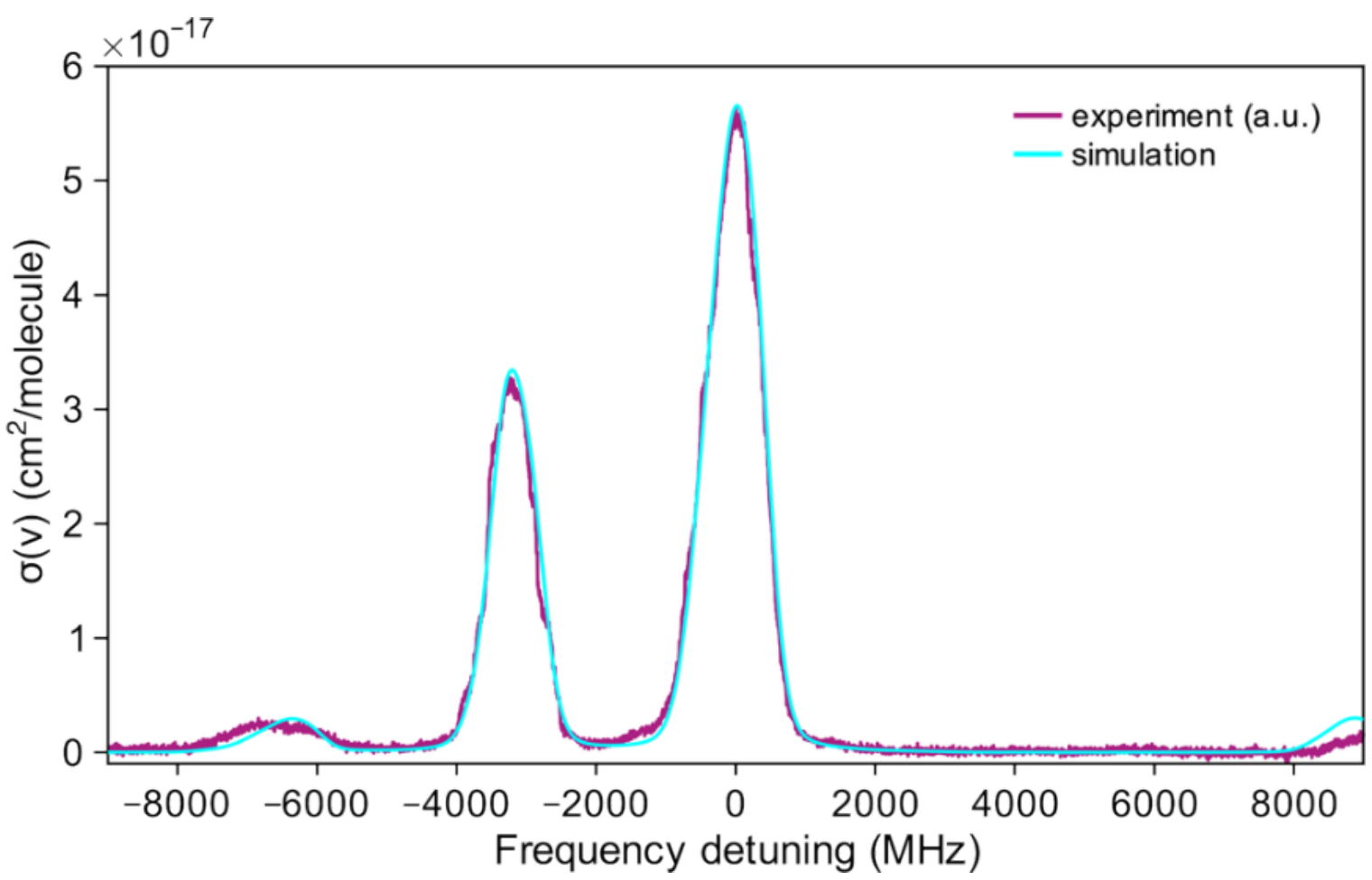

**Figure S4.** Experimental absorption spectrum of $I_2$ in a 90 sccm flow of He in the extracavity vapor flow cell plotted with arbitrary intensity (magenta). The rovibronic $I_2$ model implemented in PGOPHER (cyan) agrees with the experimental spectrum using a Gaussian broadening of 435 MHz (0.0145 $cm^{-1}$) fwhm and Lorentzian broadening of 95.6 MHz (0.00319 $cm^{-1}$) fwhm. The frequency axis is referenced with the center of the $I_2$ absorption band at 18788.4405 $cm^{-1}$ corresponding to 0 MHz.

## S3. Additional figures

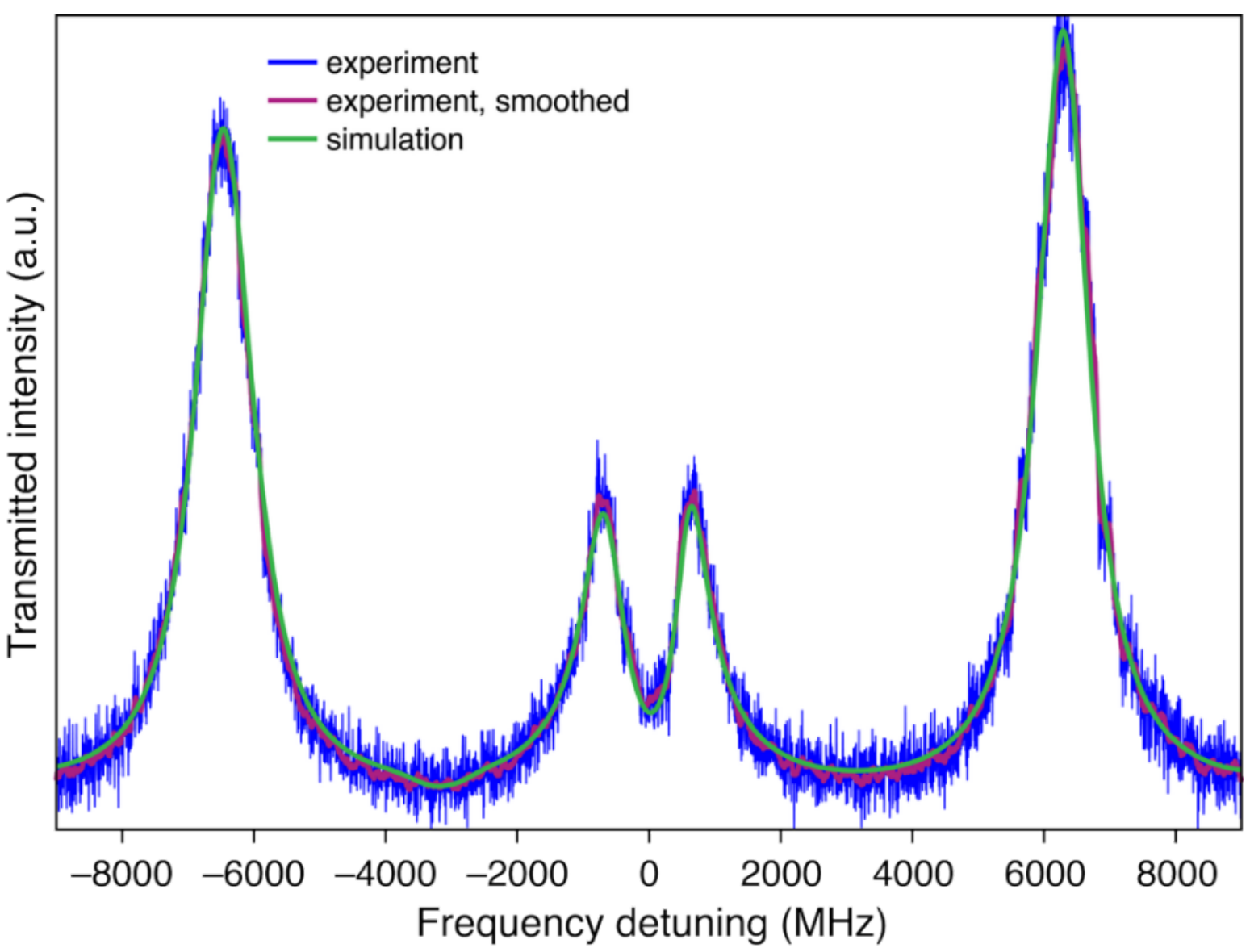

**Figure S5.** Raw experimental transmission spectrum of cavity containing $I_2$ carried in a 30 sccm flow of He (blue). The spectrum is smoothed with a moving average filter and smoothing parameter of 20 points, corresponding to a 42 MHz smoothing window (magenta). Here, the simulated cavity transmission spectrum (green) is produced by fitting eq 1 to the experimental data with an intracavity number density of $[I_2] = 4.3 \times 10^{15}$ $cm^{-3}$. The frequency axis is referenced with the center of the $I_2$ absorption band at 18788.4405 $cm^{-1}$ corresponding to 0 MHz.

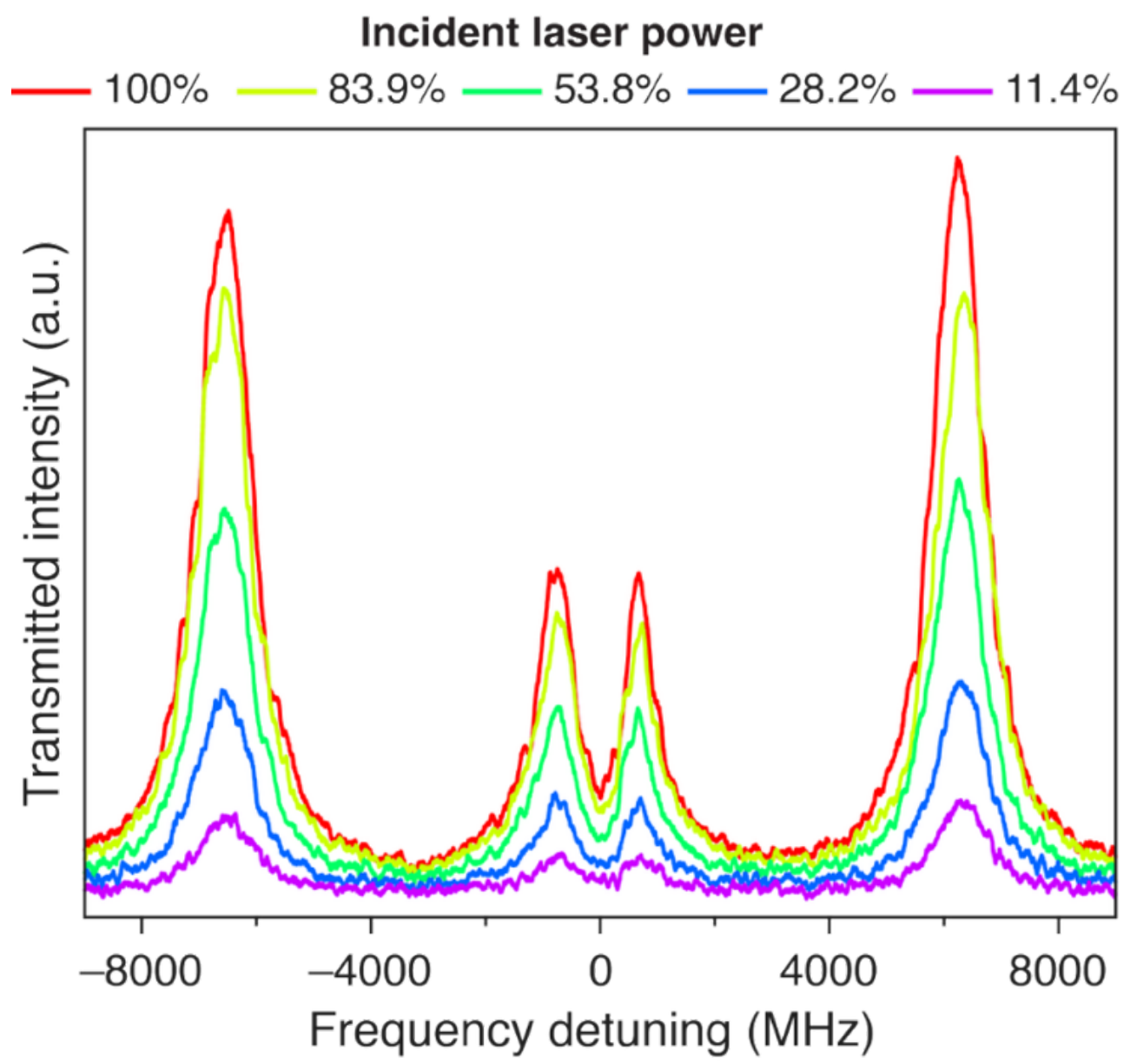

**Figure S6.** Experimental transmission spectra of cavity containing $I_2$ carried in a 30 sccm flow of He as a function of incident 532 nm laser power. 100% laser power corresponds to 31.6 μW incident on the input window of the cavity cell. As we decrease the laser power, the transmitted signal correspondingly drops in intensity, but the peak positions remain consistent. In each trace, the cavity fringe is locked on resonance with the $I_2$ transition at 18788.4405 $cm^{-1}$. The frequency axis is referenced with this resonance corresponding to 0 MHz.

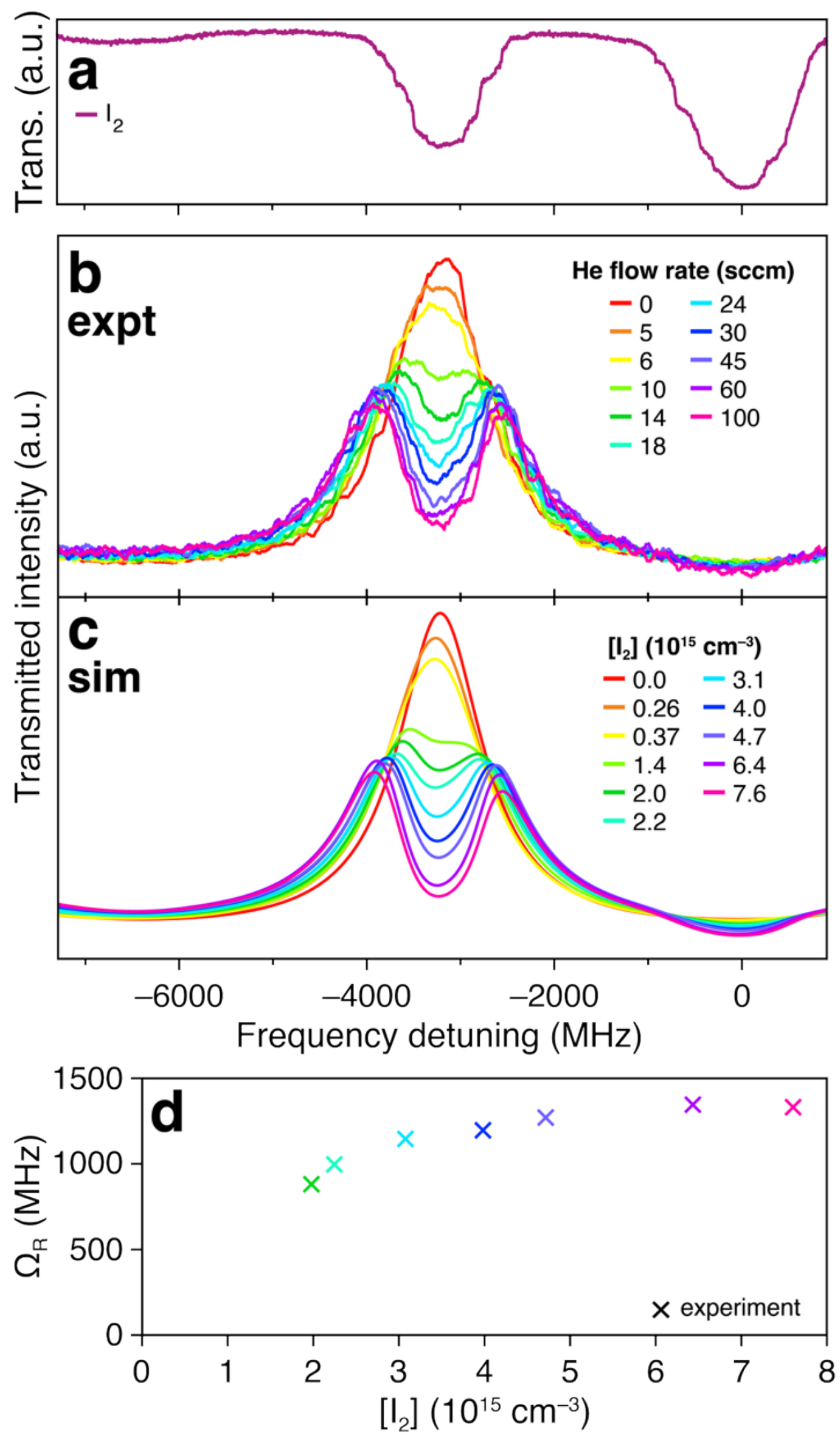

**Figure S7. (a)** Transmission spectrum of extracavity vapor flow cell containing $I_2$ carried in a 90 sccm flow of He. The frequency axis is referenced with the center of the primary absorption band at 18788.4405 cm$^{-1}$ corresponding to 0 MHz. **(b)** Smoothed experimental transmission spectra of the cavity under systematic tuning of the intracavity $I_2$ number density via the He carrier gas flow rate. **(c)** Simulated cavity transmission spectra obtained via fitting of the corresponding experimental traces with the Fabry-Pérot cavity transmission expression in eq 1. **(d)** Peak splittings extracted from the experimental spectra are plotted with color-coding to match panels (b) and (c) as a function of $[I_2]$. Here, the Rabi splitting does not ultimately exceed the cavity linewidth, and the system therefore does not quite reach the strong coupling regime. Over the course of these experiments the fitted mirror reflectivity drops from 76.2 to 71.1% corresponding to a cavity mode linewidth increase from 1093 to 1374 MHz.